# Experimental evidence and control of the bulk-mediated intersurface coupling in topological insulator $Bi_2Te_2Se$ nanoribbons

Zhaoguo Li[1†], Ion Garate[2†], Jian Pan[3], Xiangang Wan[1], Taishi Chen[1], Wei Ning[4], Xiaoou Zhang[1], Fengqi Song[1*], Yuze Meng[1], Xiaochen Hong[3], Xuefeng Wang[5], Li Pi[4], Xinran Wang[5], Baigeng Wang[1*], Shiyan Li[3], Leonid Glazman[6] and Guanghou Wang[1]

[1] National Laboratory of Solid State Microstructures, Collaborative Innovation Center of Advanced Microstructures, and Department of Physics, Nanjing University, Nanjing 210093, P. R. China

[2] Département de Physique and Regroupement Québécois sur les Matériaux de Pointe, Université de Sherbrooke, Sherbrooke, Québec, Canada J1K 2R1

[3] State Key Laboratory of Surface Physics, Department of Physics, and Laboratory of Advanced Materials, Fudan University, Shanghai 200433, P. R. China

[4] High Magnetic Field Laboratory, Chinese Academy of Sciences, Hefei 230031, Anhui, P. R. China

[5] School of Electronic Science and Engineering, Nanjing University, Nanjing 210093, P. R. China

[6] Department of Physics, Yale University, New Haven, Connecticut 06520, USA

---

[†]These authors contributed equally.

[*]Authors to whom the correspondence should be addressed: F. S. (songfengqi@nju.edu.cn) and B. W. (bgwang@nju.edu.cn). Fax: +86-25-83595535.

**Abstract:**

Nearly a decade after the discovery of topological insulators (TIs), the important task of identifying and characterizing their topological surface states through electrical transport experiments remains incomplete. The interpretation of these experiments is made difficult by the presence of residual bulk carriers and their coupling to surface states, which is not yet well understood. In this work, we present the first evidence for the existence and control of bulk-surface coupling in $Bi_2Te_2Se$ nanoribbons, which are promising platforms for future TI-based devices. Our magnetoresistance measurements reveal that the number of coherent channels contributing to quantum interference in the nanoribbons changes abruptly when the film thickness exceeds the bulk phase relaxation length. We interpret this observation as an evidence for bulk-mediated coupling between metallic states located on opposite surfaces. This hypothesis is supported by additional magnetoresistance measurements conducted under a set of gate voltages and in a parallel magnetic field, the latter of which alters the intersurface coupling in a controllable way.

**Introduction**

The discovery of topological insulators[1-4] (TIs) has ignited a race to develop novel quantum devices aimed at exploiting the peculiar transport and magnetoelectric properties of these materials[5-8]. Most prototype devices require TIs with (i) a perfectly insulating bulk and (ii) metallic and independent topological surface states (TSS). However, these desiderata remain difficult to satisfy in practice, due to unintended bulk doping and due to bulk-surface coupling[9] (BSC).

Ultrathin films of thickness ≲10 nm are optimal platforms to achieve bulk insulation because a gate voltage can tune the chemical potential inside the bulk bandgap across the entire film[10]. Yet, TSS in these films are not independent: the rate of direct intersurface electron tunneling often exceeds the phase relaxation rate and consequently the TSS localized on opposite surfaces merge into a single conduction channel without topological protection[11-14].

In thicker films, where direct intersurface tunneling is exponentially suppressed, BSC makes it possible that bulk carriers mediate an indirect communication between TSS localized on opposite surfaces, resulting once again in topologically trivial transport[15,16]. If the bulk is metallic, this bulk-mediated intersurface coupling can be strong even in films that are hundreds of nanometers thick[17-19]. The difficulty to achieve independent TSS channels in ultrathin and thicker films alike casts uncertainty on the practical potential of TI-based devices.

Prospects appear to be brighter for recently synthesized[20] $Bi_2Te_2Se$, where the chemical potential lies inside the bulk gap without having to resort to neither ultrathin films nor gate voltages. Yet, although this material possesses relatively few bulk carriers, it is not known whether or not these residual bulk carriers can significantly couple the TSS localized on different surfaces. Because any bulk-mediated communication between different TSS can pose an obstacle for practical applications, the understanding and control of this coupling is an essential step towards making TI-based quantum devices.

The objective of the present work is not only to understand the magnitude and origin of the intersurface coupling in $Bi_2Te_2Se$ films, but also to find ways to control it. To that end, we perform systematic magnetoresistance (MR) measurements for eighteen samples of different thicknesses. We find that the intersurface coupling is strongly reduced when the sample thickness exceeds the bulk phase relaxation length. In films that are thinner than the bulk phase relaxation length, we observe that a magnetic field applied parallel to the film can gradually decouple the two surfaces. These findings suggest that bulk carriers mediate the intersurface coupling. In addition, the magnetic-field control of the coupling, established here for the first time, may be useful for future TI-based devices.

**Intersurface coupling-decoupling transition revealed by weak antilocalization**

We have analyzed the transport characteristics of eighteen $Bi_2Te_2Se$ nanoribbons, whose thicknesses ($H$) range from $H = 30$ nm to $H = 100$ nm. All samples are exfoliated from the same crystal and grown by a high-temperature sintering method[21]. Their preparation and parameters are detailed in the Method and Supplementary

Material. At high temperatures, the resistivity of a typical sample shows a thermally activated behavior[20] with a gap of 1-10meV[cf. Fig.1(a) and the Supplementary Material]. At low temperatures, the bulk resistivity often saturates and yields a low mobility of about 100 cm$^2$/Vs. Although suppressed, this mobility likely falls within the diffusive (metallic) regime[22].

In Fig.1, we illustrate the MR curves of a typical sample (S9), a 98 nm-thick and 880 nm-wide ribbon, for different directions of the applied magnetic field. Figure 1(b) shows the perpendicular-field MR at different temperatures. The zero-field dip in MR becomes shallower and oscillations disappear with increasing temperatures, which confirms the quantum nature of the MR[23]. These features are attributed to weak antilocalization[24] (WAL) and universal conductance fluctuations[21]. Hereafter we focus on the WAL features.

Figure 1(c) shows the MR curves for $\theta = 0°$ and 90° respectively, where $\theta$ is the angle between the field and the unit vector normal to the film. Recognizing that the surface states (SSs) are largely insensitive to in-plane fields, we ascribe the MR contribution at $\theta = 90°$ to bulk carriers. In doing so, we neglect the SS contribution to the parallel-field MR, which is nonzero due to the BSC. Likewise, we neglect the dependence of the bulk contribution on the field direction, which arises due to the BSC as well as due to the finite thickness of the ribbons. These approximations are likely justified because our nanoribbons are not thin compared to the bulk phase relaxation length[25].

In Fig. 1(d), we have subtracted the bulk ($\theta = 90°$) contribution from the magnetoconductance (MC) obtained at other angles[26]. Upon this subtraction, all MC curves coincide with each other when plotted as a function of the perpendicular component of the magnetic field. The two-dimensional (2D) nature of the WAL is thus demonstrated[27]. Accordingly, we fit the MC curves of Fig. 1(d) to the Hikami-Larkin-Nagaoka (HLN) formula[28]

$$\Delta G_\square(B) = \alpha \frac{e^2}{2\pi^2 \hbar} \left[ \ln\left(\frac{\hbar}{4eL_{\phi,SS}^2 B}\right) - \psi\left(\frac{1}{2} + \frac{\hbar}{4eL_{\phi,SS}^2 B}\right) \right] , \quad (1)$$

where $\Delta G_\square(B) = G_\square(B) - G_\square(0)$ and $G_\square = G \cdot (L/W)$. Here, $G = 1/R$ is the conductance of the ribbon, $R$ is its resistance, $W$ is the ribbon width and $L$ is its length. Also, $L_{\phi,SS}$ is the phase relaxation length of SSs, $\psi(x)$ is the digamma function and α is a coefficient that reflects the number of independent conduction channels on the surfaces of the film[9,14]. For sample S9, the best fit yields α = 0.28 and a surface phase relaxation length of $L_{\phi,SS}$ = 141 nm. The bulk dephasing length of $L_{\phi,B}$ = 66 nm is also obtained by analyzing the MR at θ = 90° (cf. Supplementary Material, Section 3).

Figure 2(a) shows the phase relaxation length as a function of sample thickness across different samples. All the values of $L_{\phi,B}$ are scattered around 60 nm and are much smaller than $L_{\phi,SS}$. Fig. 2(b) shows that α ≈ 0.5 when $H < L_{\phi,B}$ and α ≈ 0.25 when $H > L_{\phi,B}$. Because the bulk contribution has already been subtracted, the abrupt thickness-dependent change in α is interpreted as a change in the number of independent surface conduction channels. The value of α is affected by any phase-coherent coupling that may exist between them. A single, isolated TSS leads to α = 0.5 (WAL) regardless of the band parameters. In contrast, the contribution from a single and isolated trivial 2D electron gas (2DEG) can range between α = −1 [weak localization (WL)] for weak spin-orbit coupling and α=0.5 (WAL) for strong spin-orbit coupling. The fact that α ≈ 0.5 for the thinnest films suggests a strong and phase-coherent intersurface coupling therein. In addition, the observation of α < 0.5 for the thicker films reveals the existence of at least one topologically-trivial surface 2DEG. This extra channel must be in the WL (rather than WAL) regime, which is possible if its states at the Fermi surface are weakly spin-orbit coupled[9,29]. Finally, measuring α < 0.5 in thicker films means that the coupling between the TSS and a trivial 2DEG is weak. Since the transition in the value of α takes place when the film thickness is approximately equal to the bulk phase relaxation length, we propose that its origin is bulk-mediated intersurface coupling. Next, we demonstrate that the trivial 2DEG is located on the top surface of the ribbon.

**Evidence for the existence of a trivial 2DEG on the top surface**

Figure 3(a) shows the measured resistance as a function of the back-gate-voltage ($V_G$) for a 60 nm-thick sample (S17). The MC curves at various $V_G$ are displayed in Fig. 3(b). Fitting these curves to the HLN equation, we obtain α and $L_{\phi,SS}$ as functions of $V_G$. We observe that $L_{\phi,SS}$ varies with $V_G$ [Fig. 3(d)], while α does not [Fig. 3(c)].

The change in $L_{\phi,SS}$ comes from the bottom surface because the Fermi level of the top SS is unaffected by the gate voltage (due to the large film thickness). As $V_G$ is made more negative, the Fermi level ($E_F$) of the bottom surface approaches the Dirac point, the carrier density is reduced and the Coulomb screening weakens, thereby resulting in an enhancement of electron-electron scattering rate[30]. This explains the observed resistance maximumin[Fig. 3(e)] as well as the decrease of $L_{\phi,SS}$. A similar observation has been made in other TIs[15,16,29] and graphene[31].

The independence of α from $V_G$ provides negative evidence for having an independent topologically trivial 2DEG on the bottom surface. The transport signature of such 2DEG would be an abrupt change[29] in α as the negative gate voltage moves the Fermi energy across the band edge of the 2DEG. This is in contradiction to our observations. Thus, on the bottom surface $E_F$ appears to intersectonly with the TSS for the fullrange of $V_G$.

In the previous section,we inferred the existence of a topologically trivial 2DEG based on the observation of α < 0.5 and based on the analysis of quantum oscillations. In view of the back-gate measurements, we conclude that the topologically trivial 2DEG must be located on the *top* surface. This is reasonable because the top surface is exposed to air, which can lead to a strong enough band bending to bind a 2DEG at the surface[29,32-34].

In sum, the Fermi level intersects both the TSS and a trivial 2DEG on the top surface, while it intersects only the TSS on the bottom surface. The thickness-dependent change in α is due to the coupling-decoupling transition between the hybrid 2DEG+TSS state localized at the top surface and the TSS localized at the bottom surface. In the strong intersurface coupling regime, the spin-singlet Cooperons of the two TSS combine to yield α ≈ 0.5, while the spin-triplet Cooperon from the trivial 2DEG is gapped. In the weak intersurface coupling regime, the 2DEG contribution drives α < 0.5, provided that the spin-orbit interaction therein is small and provided that it is not strongly coupled to the TSS localized on the same surface.

**The physics of the bulk-mediated intersurface coupling**

One thickness-dependent mechanism that can couple the top and bottom surfaces is direct electron tunneling[10], which has been invoked to interpret numerous coupling-decoupling transitions[35-42] ENREF 35. Direct tunneling is the only way to coherently couple the surfaces when there are no bulk states at the Fermi level. In our case, this mechanism plays no role because the Fermi energy intersects metallic bulk states and because the sample thickness far exceeds the SS penetration depth into the bulk.

Here we address another way to couple the two surfaces through conducting bulk states[9]. Electrons from one surface can tunnel to the bulk, diffuse toward the vicinity of the opposite surface, and tunnel onto it. This process couples the two surfaces into a single coherent channel by bulk electrons provided that

$$2\tau_{SB} + H^2/D_B < \tau_{\phi,SS} \qquad (2a)$$

$$H^2/D_B < \tau_{\phi,B}. \qquad (2b)$$

Here, $\tau_{\phi,B(SS)} = (L_{\phi,B(SS)})^2/D_{B(SS)}$ is the bulk (surface) phase relaxation time, $D_{B(SS)}$ is the bulk (surface) diffusion constant, $\tau_{SB}^{-1}$ is the tunneling rate between the SS and the bulk states and $H^2/D_B$ is the time it takes for an electron to diffuse across a film of thickness $H$. The factor of 2 in front of $\tau_{SB}$ comes from the unessential assumption that the BSC is the same on both surfaces. Equation (2) does not incorporate inelastic (e.g. phonon-induced) hopping processes across the film. Although these processes contribute to the electronic transport, they are not phase coherent and therefore do not alter the value of α.

As indicated by Eq. (2), the bulk-mediated coupling between the surfaces depends not only on the bulk-surface tunneling rate, but also on the intersurface diffusion time. In thin samples with strong bulk-surface tunneling or with high bulk mobility, electrons diffusing from a surface to the other keep their phase coherence. As a result, the two surfaces contribute to MR as a single channel. When $H$ or $\tau_{SB}$ increases enough so that $2\tau_{SB} + H^2/D_B > \tau_{\phi,SS}$ or $H^2/D_B > \tau_{\phi,B}$, then the two surfaces are decoupled and the value of α changes. In previous studies[15-19], the BSC was considered only in samples with $H \ll L_{\phi,B}$ (i.e. negligible $H^2/D_B$), where the bulk-mediated coupling between the surfaces depends only on $\tau_{SB}/\tau_{\phi,SS}$. Equation (2) generalizes the previous considerations after taking into account the propagation time of electrons across the film.

The observed change of α when $H \approx L_{\phi,B}$ can be explained with Eq. (2) as long as $\tau_{SB} \ll H^2/D_B$.[43] Clearly, when the film thickness exceeds the bulk phase relaxation length, bulk carriers are unable to coherently couple the two TSS. To our knowledge, there are no previous studies demonstrating that the intersurface coupling depends on the phase coherence of bulk carriers.

**Magnetic-field control of the intersurface coupling**

The bulk-mediated intersurface coupling depends crucially on the quantum coherence of electrons as they travel from one surface to another. Consequently, external perturbations that reduce bulk coherence tend to decouple the two surfaces.

Here we measure the influence of a parallel magnetic field on α. The motivation for applying a parallel field is that it reduces bulk coherence without altering the properties of the SSs[44].

In presence of a parallel field ($B_\parallel$), the effective bulk phase relaxation time is reduced[45] via $\tau_{\phi,B}^{-1}(B_\parallel) = \tau_{\phi,B}^{-1}(0) + \frac{H^2}{6l_B^2}\frac{eD_B B_\parallel}{\hbar}$, where $l_B = \sqrt{\hbar/eB_\parallel}$ is the magnetic length. In addition (cf. Supplementary Material, Section6) the effective bulk-surface scattering rate is renormalized via $\tau_{SB,\text{eff}}^{-1}(B_\parallel) = \tau_{SB}^{-1}\exp\left(-\frac{H^2}{8l_B^2}\right)$. Then, the condition for the two surfaces to couple into a single coherent channel is

$$2\tau_{SB,\text{eff}}(B_\parallel) + H^2/D_B < \tau_{\phi,SS} \tag{3a}$$

$$H^2/D_B < \tau_{\phi,B}(B_\parallel). \tag{3b}$$

Because $B_\parallel$ makes $\tau_{SB,\text{eff}}(B_\parallel)$ longer and $\tau_{\phi,B}(B_\parallel)$ shorter, a parallel field degrades intersurface coupling. This effect becomes significant when $l_B < H$.

In order to find the dependence of α on $B_\parallel$, we measure $\Delta G_{SS}(B_\perp, B_\parallel) = G(B_\perp, B_\parallel) - G\left(0, \sqrt{B_\perp^2 + B_\parallel^2}\right)$ as a function of $B_\perp$, for a fixed value of $B_\parallel$ [inset of Fig. 4(b)]. Here, $B_\perp$ is a field perpendicular to the film. Subtracting $G\left(0, \sqrt{B_\perp^2 + B_\parallel^2}\right)$ removes the contribution from bulk states. We fit $\Delta G_{SS}(B_\perp, B_\parallel)$ to the HLN equation and extract $L_{\phi,SS}$ and α for that particular $B_\parallel$. By repeating the measurement for different values of $B_\parallel$, we establish the dependences of $L_{\phi,SS}$ and α on $B_\parallel$.

The results for a 45 nm-thick sample (S16) are shown in Fig. 4. When $B_\parallel = 0$, α ≈ 0.5 because $H < L_{\phi,B}$. As $B_\parallel$ grows from 0 T to 1 T, α decays to 0.38. This trend is in qualitative agreement with Eq. (3). When $B_\parallel$ = 1 T, the magnetic length is about 26 nm, which results in $\tau_{SB,\text{eff}}^{-1}(B_\parallel) \approx 0.7\tau_{SB}^{-1}$ and $L_{\phi,B}(B_\parallel) \approx 30$ nm. The main effect of $B_\parallel$ in the intersurface coupling is thus to decrease the effective bulk phase relaxation length, which becomes smaller than the film thickness as $B_\parallel$ increases. For completeness, Fig. 4(b) shows $L_{\phi,SS}$ as a function of $B_\parallel$. A perfectly in-plane field should not change of the surface phase relaxation rate. We speculate that the slight dependence we observe is due to $B_\parallel$ not being perfectly parallel to the sample surface.

Overall, the decay of α with $B_\parallel$ supports the hypothesis that the intersurface coupling in our nanoribbons is mediated by bulk states. To our knowledge, the control of intersurface coupling by means of a parallel magnetic field has not been previously achieved.

**Conclusions**

We have measured the quantum corrections to conductivity in $Bi_2Te_2Se$ nanoribbons. The effective number of surface conduction channels rises as the thickness of the sample exceeds the bulk phase relaxation length. The origin of this increase appears to be the decoupling between conducting states localized on opposite surfaces. We propose that bulk carriers mediate the intersurface coupling. These carriers lose phase coherence over a distance of a phase relaxation length; as such, they are unable to coherently couple the two surfaces when the thickness of the sample exceeds the phase relaxation length. For a sample of a fixed thickness, a parallel magnetic field (which effectively reduces bulk coherence without altering the SSs) leads to the gradual decoupling between the surfaces. Overall, our study provides new insight on the origin and control of bulk-mediated intersurface coupling in topological insulating devices.

**Methods**

$Bi_2Te_2Se$ single crystals were grown by melting high purity Bi (5N), Te (5N) and Se (5N) powders in sealed quartz ampoules. Using Scotch tape, $Bi_2Te_2Se$ nanoribbons were exfoliated and deposited on doped Si covered with 300 nm $SiO_2$. Then, Au electrodes were applied onto the nanoribbons by a standard lift-off procedure. $Bi_2Te_2Se$ nanoribbons with thicknesses ranging from 30 to 100 nm were identified by Atomic Force Microscopy (AFM). The resistance of all samples was measured with a four-probe configuration. The magnetotransport measurements were carried out in the Quantum Design Physical Property Measurement System. The gate tuning experiment was carried on a homemade measurement system, which contains a 5-Tesla superconductor magnet, a 4 K G-M refrigerator and a Keithley 4200-SCS as the electrical characterization instrument. The parallel-field tuning experiment is performed on an Oxford vector rotate magnet system. More details can be found in the Supplementary Section.

**Acknowledgements**

We thank the National Key Projects for Basic Research of China (Grant Nos. 2013CB922103, 2011CB922103 and 2010CB923401), the National Natural Science Foundation of China (Grant Nos. 11023002, 11134005, 60825402 and 61176088), NSF of Jiangsu province (Nos. BK2011592, BK20130016, BK20130054,

BK2012322 and BK2012302), the PAPD project and the Fundamental Research Funds for the Central Universities, the Program for the New Century Excellent Talents in University for financially supporting this work. Helpful assistance from the Nanofabrication and Characterization Center at Physics College of Nanjing University and Prof. Yuheng Zhang of High Magnetic Field Laboratory CAS are acknowledged. We would like to thank Prof. Mark Reed of Yale University, Feng Miao of Nanjing University and Yongqing Li of Institute of Physics CAS for meaningful discussions. I.G. acknowledges financial support from Université de Sherbrooke and from Canada's National Science and Engineering Research Council. Work at Yale University is supported by NSF DMR Grant 1206612.

than the surface phase relaxation rate [O. E. Raichev & P. Vasilopoulos, J. Phys.: Condens. Matter 12, 589 (2000)]. However, the large thicknesses of our films rules out this possibility. Another way to obtain parallel-field MR without invoking bulk states is through the effect of the magnetic field on the side surfaces. However, this mechanism does not apply in our samples because the in-plane component of the applied magnetic field is parallel to the current.

43   However, the fact that α is independent of the gate voltage seems puzzling. The band bending due to a negative gate voltage induces a depletion layer between the bottom SS and the bulk carriers. This should in principle result in a large increase of τSB at the bottom surface, thereby decoupling the two surfaces. In fact, in metallic systems it has been confirmed that a gate voltage can break the intersurface coupling (Refs. 15, 16, 19). A possible way to reconcile bulk-mediated intersurface coupling with a gate-independent value of α is to invoke localized impurity states under the depletion barrier. Although these states do not contribute to conductivity, they facilitate electron tunneling from the surface to the bulk. Under resonant conditions, the bulk-surface tunneling probability approaches unity regardless of the height of the depletion barrier [L. I. Glazman & R. I. Shekhter, Sov. Phys. JETP 67, 163 (1988)]. This may then explain why τSB remains small regardless of the gate voltage.

44  Increasing the temperature is an alternative way to break the conditions in Eq. (2). However, increasing the temperature reduces both surface an bulk phase relaxation lengths, thus making it less straight forward to test the idea of bulk-mediated intersurface coupling.

45  Al'tshuler, B. L. & Aronov, A. G. Magnetoresistance of thin films and of wires in a longitudinal magnetic field. *JETP Lett.* **33**, 499-501 (1981).

**Figure captions:**

**Figure 1. Magnetotransport in a $Bi_2Te_2Se$ nanoribbon with a thickness of 98 nm.** (a) Temperature-dependence of the resistance. The left inset shows the AFM image of the sample and the right inset shows a schematic diagram of the measurement configuration. (b) Low-field magnetoresistance (MR) at temperatures $T$=2, 6, 10, 15 and 25K. The positive MR is a consequence of weak antilocalization (WAL). (c) The MR curves at $\theta = 0°$ and $90°$, with WAL at low fields and universal conductance fluctuations at high fields. (d) The sheet conductance ($\Delta G_\square$) as a function of the perpendicular component of the magnetic field, measured at $T = 2$ K for various angles $\theta$. All the curves coincide with each other. The solid curve is a fit to Eq. (1).

**Figure 2. Relating the thickness-dependent α to the quantum coherence of the bulk electrons.** (a) Phase relaxation length of the surface channels ($L_{\phi,SS}$, squares) and the bulk channel ($L_{\phi,B}$, circles) as a function of the sample thickness ($H$). The solid line represents $L_\phi = H$. The hatched area covers the range of $L_{\phi,B}$ across different samples. (b) α as a function of $H$ across different samples. The dashed horizontal line corresponds to α = 1/2. The dashed vertical line in (a) and (b) corresponds to $H$ = 62 nm.

**Figure 3. Influence of a back gate on transport**, **at T=4.6 K**. (a) Resistance as a function of the voltage ($V_G$) applied to the back gate. (b) Low-field magnetoconductance for different values of $V_G$. The solid curves are the best fits to the HLN equation. For clarity, the curves are shifted vertically with respect to one another. (c) $V_G$ − dependence of α. (d) $V_G$− dependence of the surface phase relaxation

length $L_{\phi,SS}$. The dashed curves in (c) and (d) are guides for the eyes. (e) Tentative band diagram describing our samples. The Fermi level ($E_F$) intersects both the TSS and the 2DEG on the top surface, and the TSS on the bottom surface. A back gate displaces the Fermi energy near the bottom surface.

**Figure 4. Control of intersurface coupling by a parallel magnetic field, at T=1.5 K.** (a) $\Delta G(B_\perp)$ curves for different parallel fields ($B_\parallel$). The bulk contribution is subtracted. The solid curves are the best fits to the HLN equation. (b) $B_\parallel$- dependence of the surface phase relaxation length $L_{\phi,SS}$. The slight decrease of $L_{\phi,SS}$ with increasing $B_\parallel$ may be due to the fact that $B_\parallel$ is not perfectly parallel to the sample's surface. The inset shows a schematic diagram of the field configuration. (c) $B_\parallel$-dependence of α. As $B_\parallel$ is increased, the phase coherence of bulk electrons gets reduced and thus the bulk-mediated intersurface coupling is degraded. This is consistent with the observed reduction in α.